\documentclass[journal,twoside]{IEEEtran}
\usepackage{xcolor}
\usepackage{graphicx}
\usepackage[colorlinks, citecolor=blue]{hyperref}
\usepackage{array}
\usepackage{cite}
\usepackage{amsmath,amssymb,amsfonts}
\usepackage{algorithmic}
\usepackage{graphicx}
\usepackage{textcomp}
\usepackage{multirow}
\usepackage{mathtools}
\usepackage{soul}
\usepackage{subfigure}

\usepackage[T1]{fontenc}%
\usepackage[utf8]{inputenc}%
\usepackage{lmodern}%
\usepackage{textcomp}%
\usepackage{lastpage}%
\usepackage{graphicx}%

\newcolumntype{C}[1]{>{\centering\let\newline\\\arraybackslash\hspace{0pt}}m{#1}}
\newcolumntype{L}[1]{>{\raggedright\let\newline\\\arraybackslash\hspace{0pt}}m{#1}}

\usepackage{etoolbox}
\makeatletter
\patchcmd{\@makecaption}
  {\scshape}
  {}
  {}
  {}
\makeatletter
\patchcmd{\@makecaption}
  {\\}
  {.\ }
  {}
  {}
\makeatother

\begin{document}

\title{\huge{SeaPlace: Process Variation Aware Placement for Reliable Combinational Circuits against SETs and METs}}
%
%
%

\author{Kiarash Saremi, Hossein Pedram, Behnam Ghavami, Mohsen Raji, Zhenman Fang, Lesley Shannon
\thanks{K. Saremi and H. Pedram are with the Department of Computer Engineering, Amirkabir University of Technology, Iran (e-mails: \{k.saremi, pedram\}@aut.ac.ir)}
\thanks{M. Raji is with the School of Electrical and Computer Engineering, Shiraz University, Iran (e-mail: mraji@shirazu.ac.ir)}
\thanks{B. Ghavami is with the Shahid Bahonar University of Kerman, Iran and Simon Fraser University, Canada (e-mail: ghavami@uk.ac.ir)}
\thanks{Z. Fang and L. Shannon are with School of Engineering Science, Simon Fraser University, Canada, (e-mails: \{zhenman, lshannon\}@sfu.ca)}
}

\markboth{}%
{Saremi \MakeLowercase{\textit{et al.}:} SeaPlace: Process Variation Aware Placement for Reliable Combinational Circuits}

\maketitle
\begin{abstract}
Nowadays nanoscale combinational circuits are facing significant reliability challenges including soft errors and process variations. This paper presents novel process variation-aware placement strategies that include two algorithms to increase the reliability of combinational circuits against both Single Event Transients (SETs) and Multiple Event Transients (METs).  The first proposed algorithm is a global placement method (called SeaPlace-G) that places the cells for hardening the circuit against SETs by solving a quadratic formulation. Afterwards, a detailed placement algorithm (named SeaPlace-D) is proposed to increase the circuit reliability against METs by solving a linear programming optimization problem. Experimental results show that SeaPlace-G and SeaPlace-D averagely achieve 41.78\% and 32.04\% soft error reliability improvement against SET and MET, respectively. Moreover, when SeaPlace-D is followed by SeaPlace-G, MET reduction can be improved by up to 53.3\%.

\end{abstract}


\section{Introduction}

\IEEEPARstart{T}{he} reliability of nanoscale digital circuits is becoming increasingly important in recent years \cite{1}. Due to aggressive technology scaling, the vulnerability of digital circuits to single-event transients (SETs) and single-event upsets (SEUs) caused by radiation induced transient faults is increasing \cite{1,2,3}. On the other hand, process variations caused by technology scaling has introduced new challenges in the circuit reliability \cite{4}. As shown in \cite{5}, process variations can impose 10\%-50\% variations to the soft error rate (SER) of circuits, leading to significant unreliable fabricated circuits. Hence, it is critical to address these issues to improve reliability of digital circuit designs.

SETs and SEUs are transient faults threatening the soft error reliability of circuits and are generated when a high energy particle strikes in combinational logics and sequential elements of circuits, respectively \cite{6}. Due to technology scaling, the critical charge has reduced and a high-energy particle can affect several adjacent cells in a circuit causing Multiple Bit Upset (MBUs) and Multiple Event Transients (METs) in memory elements and combinational circuits, respectively \cite{7}. When a transient fault is propagated through the combinational component and latched into a sequential element, a soft error is happened. Its occurrence rate is evaluated by Soft Error Rate (SER) metric \cite{8}. The recent studies indicate that SER of the combinational component of a circuit comprises a remarkable portion of the SER of the entire circuit \cite{9,10}. Moreover, recent studies have shown that a notable fraction of particle strikes results in METs \cite{11} and for technologies below 50 nm, the probability of MET occurrence is equal to that of SETs \cite{12}. Therefore, in order to increase the circuit reliability effectively, both SETs and METs need to be considered in hardening techniques.

Over the past decade, various design techniques have been proposed to increase the soft error reliability of digital circuits. One promising approach is to use an SER-aware placement algorithm instead of a traditional placement algorithm in the physical design process. On the other hand, the process variations have a remarkable impact on SER \cite{26} because of exponential dependence of SER on process parameters \cite{28}. Process variations can be decomposed into two categories \cite{28}: within-die (WID) and die-to-die (D2D) variations. WID variations model the variations within a die and D2D variations model the variations on identical cells of a circuit on various dies. The WID variations greatly affect the vulnerability of a cell against soft errors \cite{29}, and thus SER of the circuit. And the effects of WID variations on process parameters of a cell is identified after performing the placement process (note that D2D variations affect all cells of a circuit in the same manner and hence, it does not change the vulnerable cells). Therefore, an important question naturally arises: \textit{"how important is it to consider the effects of WID variations while trying to find a SER-aware placement for a circuit?"}

In this paper, we propose SeaPlace, a WID process variation-aware and \underline{s}oft \underline{e}rror \underline{a}ware \underline{place}ment methodology. The goal of SeaPlace is to use WID variation information in placement algorithms to improve the soft error reliability of combinational circuits against both SETs and METs. To the best of our knowledge, this is the first work that models and exploits WID variation information in SER-aware placement algorithms. 

We first show that the effect of WID process variations on the circuit SER is significantly affected by the placement of circuit cells. Our experiments on some random placement algorithms reveal that neglecting WID variations in placement algorithms can easily miss the opportunity of more than 30\% SER reduction in a circuit. Based on this observation, we propose two novel variation-aware placement algorithms including a global placement called as  SeaPlace-G and a detailed placement called as SeaPlace-D, to improve the circuit reliability against both SETs and METs.  SeaPlace-G is a global placement based on quadratic programming and  SeaPlace-D is a detailed placement which uses a linear programming optimization for increasing the circuit reliability. Experimental studies on various large-scale circuits including EPFL benchmark circuits show that SeaPlace-G averagely reduces the SET-originated SER by 41.78\% and SeaPlace-D achieves 32.04\% failure probability reduction against METs, on average. Furthermore, performing SeaPlace-D followed by SeaPlace-G (SeaPlace-G→SeaPlace-D) averagely decreases the failure probability by 53.3\%.

The main contributions of this paper are as follows:
\begin{enumerate}
    \item A novel WID variation-aware placement methodology called SeaPlace to improve soft error reliability against both SETs and METs including two consistent algorithms for global and detailed placement.
    \item A fast SER-aware global placement algorithm based on quadratic programming for improving soft error reliability against SETs in the presence of WID variations.
    \item A novel SER-aware detailed placement algorithm considering WID variations for hardening against METs.
    
\end{enumerate}

The remainder of the paper is organized as follows: Section II provides an overview of the related work. Section III motivates the consideration of WID variations in placements aimed for soft error reliability improvement. 
Sections IV, V and VI explain placement methodology and the proposed algorithms SeaPlace-G and SeaPlace-D, respectively. Experimental results are presented in Section VII. Finally, the conclusion is given in Section VIII.
\section{Related Work} \label{related}

Various types of design techniques, with the aim of reducing the circuit SER, have been proposed during the recent decades. Adding hardware redundancy, gate sizing, increasing load capacitance and using error correcting codes are some of these conventional techniques \cite{13,14,15,16,17,18,19}. In recent years, soft error-aware placement process is introduced as a promising technique to decrease the circuit SER \cite{20,21,22,23,24,25,26,27,5699279,betz2000vpr,betz2012architecture}. 

For SETs (or SEUs), in \cite{5699279}, a soft error reliability aware placement algorithm is presented to increase the reliability of designs to be mapped into SRAM-based FPGAs against SEUs. It modifies the placement algorithm in the VPR tool-set \cite{betz2000vpr,betz2012architecture}  by adding reliability constraints to the cost function of simulated annealing-based placement algorithm of the tool. In \cite{20}, it is stated that the masking probability of SETs generated between nearby cells with high sensitivity is high. As a result, after computing cells sensitivity information, some placement constraints are produced as bounding commands of placement design tools to place highly sensitive cells in a close proximity. Although this method takes advantage of design automation tools, it is not performed automatically and it does not have enough flexibility. 

For METs (or MUBs), in \cite{21}, via calculation of pass/fail value for all possible pairs of the circuit, all pairs are decomposed into two sets: good pairs and bad pairs. A simulated annealing-based optimization is then accomplished for maximizing the elements of good pairs set. As shown by the result, the runtime is increased greatly with the increase in circuit size. The method is not scalable and cannot be applied to large circuits. In methods presented in \cite{22} and \cite{23}, the low-pass filtering behavior of long nets is focused and it is shown that enlarging the interconnections can result in SER reduction in combinational circuit. However, these approaches negatively affect the total wirelength and the circuit delay. In \cite{24}, a post-placement approach is used to redistribute white spaces for increasing the distance between most vulnerable adjacent cells and subsequently reducing the circuit SER. There are some works that take advantage of pulse quenching effect in which simultaneous charge collection on nearby nodes of a circuit can cause to shorten or quench a radiation induced transient pulse \cite{25}. In \cite{26}, the introduced placement approach tries to increase the number of quenching units and also make them physically closer to decrease the soft error susceptibility. In \cite{27}, the authors proposed a detailed placement in which the distance between quenching units is reduced to increase the pulse quenching effect and decrease the circuit SER.

Although these prior studies have achieved considerable soft error reliability improvement for the circuits, they have overlooked one or two major issues:

\begin{enumerate}
\item Most of them only consider SETs (\cite{20,22,23}) or METs (\cite{21,24}) and do not cover all transient faults originated by both SETs and METs. It should be noted that most of the methods proposed for hardening against METs cannot be simplified to handle SETs because having multiple error sites and using their interactions is their main assumption. As the occurrence probability of METs and SETs for technologies below 50 nm is equal \cite{12}, neglecting one of them will result in ineffective soft error reliability improvement.

\item None of the previous studies addressed the effect of process variations. 
Cell locations due to WID variations quietly affect the circuit SER \cite{28}. Therefore, considering the effects of WID variations during an SER-aware placement is of great importance. Based on our characterization results which will be presented in Section III, omitting WID variations in SER-aware placements can easily lead to inefficient SER reduction by more than 30\%.
\end{enumerate}
\begin{figure*}[hbt!]
	\begin{center}
				\includegraphics[width = 0.35\textwidth]{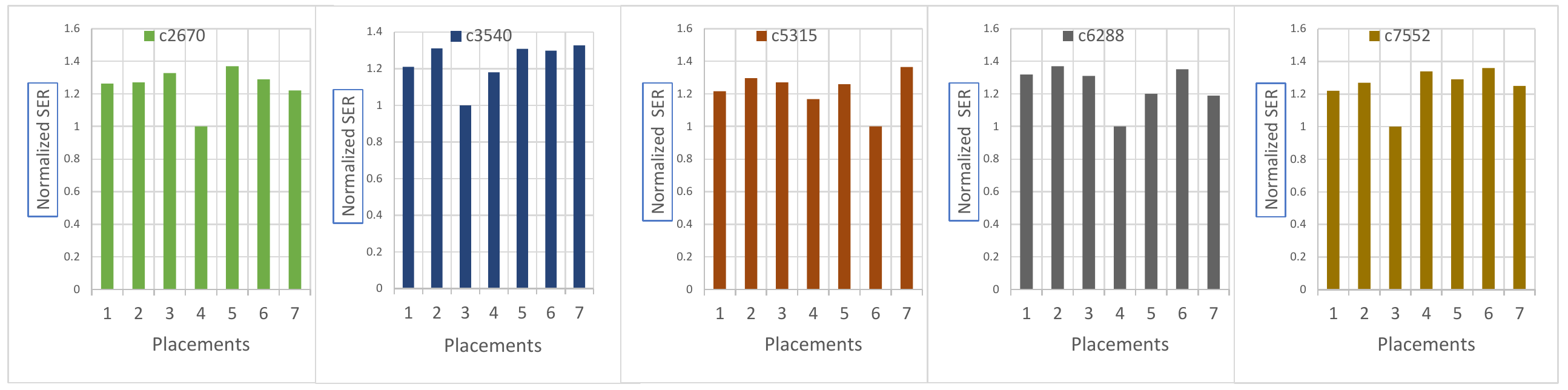}
		\vspace{-0.25in}
		\caption{Impact of WID process variations on the SER of different random placements of some ISCAS’85 circuits.}
		\vspace{-0.2in}
	\label{fig:1}
	\end{center}	
\end{figure*}

\section{Motivation: Impact of WID Process Variations on SER} \label{motivation}
WID variations impose a spatial correlation and also a random behavior to gates on different locations across the die \cite{28}, which affects gates vulnerability against generation and propagation of transient pulses generated by particle strikes \cite{29}. WID process variations can affect SER significantly: as it is shown in \cite{29}, it can result in $-33.5\%$ to $81.7\%$ variations for transistor critical charge ($Q_{critical}$). 




In this section, we characterize the effects of WID process variations during an SER-aware placement. 
For illustration purpose, we estimate the SER of a few random placements for representative combinational circuits from the ISCAS’85 benchmark suite. The technology cells are 45nm Nangate \cite{31} and the WID variations are imposed to $V_{th}$ parameter (WID variation model will be described in Section \ref{results}).

 The effects of WID variations on SER reduction is depicted in Figure~\ref{fig:1}. In this figure, for each combinational circuit of the ISCAS’85 benchmark suite, seven random placements are performed and a placement with minimum SER is considered as the base SER. Then, the SER values of the remaining placements are normalized with respect to the base SER. As shown in Figure~\ref{fig:1}, WID process variations have a great impact on SER values of different placements of a circuit, which can result in up to 36\% variations in the estimation of circuit SER. Therefore, ignoring WID process variations in SER-aware placements may lead to inaccuracy in SER reduction of combinational circuits.

\section{SeaPlace: WID Variation-Aware and Soft Error-Aware Placement} \label{proposed}
Placement is an important step during the chip physical design in which the exact geometrical location of the logic cells are determined with the aim of minimizing an objective function, usually the total wirelength. Reducing the wirelength generally increases the routability and performance of the circuit \cite{35}. In order to increase the circuit reliability, SER metrics can also be considered in the cost function of the placement.

Placement procedure typically includes two stages, i.e., global and detailed placements. Global placement distributes the cells over the placement region to optimize some objectives. The aim of global placement is to maintain a global view of placement and it focuses its attention on the cell positions globally. Because of this, global placement makes some geometrical approximations in order to simplify the placement problem and neglects some local concerns.
Detailed placement takes a placement as its input and tries to improve the solution quality. Detailed placement is more bounded than global placement, because it transforms one legal placement solution into another while optimizing the objectives \cite{36}. Detailed placement usually is performed for improving wirelength, timing and density of the layout.

In order to introduce a process variation-aware and soft error reliability-aware placement method that hardens the circuit against both SETs and METs, some considerations and limitations should be taken in advance:
 \begin{figure}[!tb]
	\begin{center}
		\includegraphics[width = 0.35\textwidth]{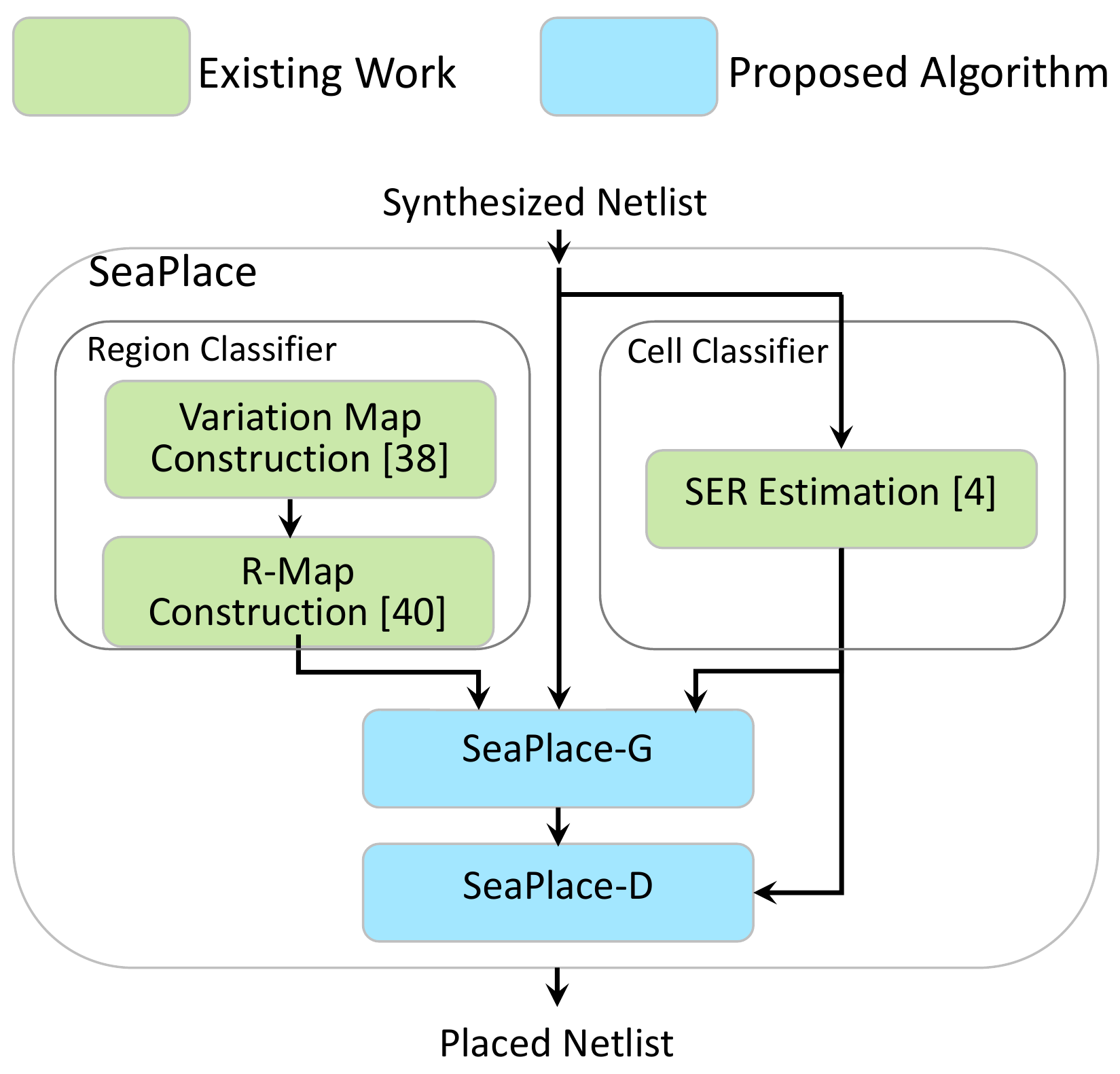}
		\vspace{-0.1in}
		\caption{Overall flow of our SeaPlace framework.}
		\vspace{-0.2in}
	\label{fig:flow}
	\end{center}	
\end{figure}

\begin{itemize}
    \item First, in order to perform hardening against METs, information of neighbor cells should be taken into account. Adding this information to the global placement step increases its complexity and runtime, as the global placement inherently does not focus on the locality and the neighbors of each cell.
    \item  Second, it is not efficient to consider both SETs and METs in the detailed placement. The reason is that, the detailed placement has not enough degree of freedom to improve vulnerability against both SETs and METs efficiently, due to the constraints applied to the detail placement.
\end{itemize}

Therefore, in this work, we consider SETs in the global placement and METs in the detail placement step. Although SETs and METs are considered in separate placement phases, mechanisms are provided such that the two phases do not negatively affect each other. 
The flow of our proposed SeaPlace framework is presented in Figure \ref{fig:flow}. This framework receives a synthesized netlist as an input and generates a soft error-aware placement. In the first phase of the framework, region classification and cell classification are performed. Then, our proposed SET-aware global placement, SeaPlace-G, is accomplished using the input netlist and the prepared information, including variation map and R-map, in the first phase. Afterwards, our MET-aware detailed placement, SeaPlace-D, is performed receiving SeaPlace-G output and some of first phase outputs (such as SER estimation results). In the following sections, we will explain our introduced flow in detail.
\section{SeaPlace-G: A Variation-Aware and SET-Aware Global Placement Algorithm} \label{segp}
\begin{figure}[!tb]
	\begin{center}
		\includegraphics[width = 0.3\textwidth]{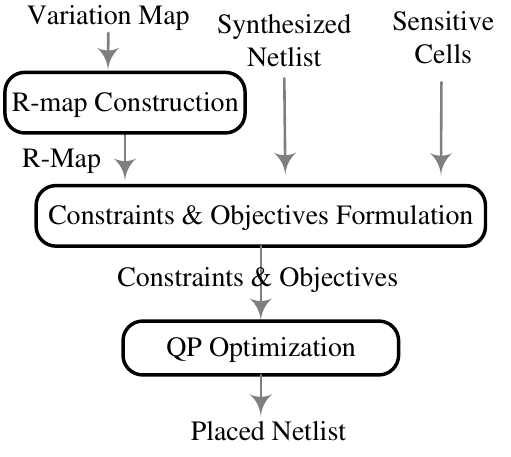}
		\vspace{-0.1in}
		\caption{Overall flow of the SeaPlace-G algorithm.}
		\vspace{-0.2in}
	\label{fig:2-2}
	\end{center}	
\end{figure}

The placement is an NP-complete problem and should be solved heuristically \cite{35}. Analytical algorithms based on Quadratic Programming (QP) optimization (also called quadratic placements) are very popular because they are quite efficient while presenting good quality of results \cite{35}. In this section, we present SeaPlace-G, a WID variation-aware and SET-aware quadratic global placement algorithm to increase soft error reliability. The overall flow of SeaPlace-G is shown in Figure \ref{fig:2-2}. The algorithm receives  an R-map extracted from variation map, a synthesized netlist and a set of sensitive cells as inputs. Constraints and objectives of the optimization are formulated in the next phase. Then, a QP optimization approach is
performed to achieve the final placement.    

\subsection{Quadratic Placement Formulation}
A Quadratic placement solves a minimization problem where its objective function is the sum of weighted squared distance of connected cells, i.e.,:
\begin{equation} \label{eq:4}
\phi = \frac{1}{2} \sum_{(a, b) \in C}^{}{w_{ab} \left[ \left(a_x - b_x \right)^2 + \left(a_y - b_y \right)^2 \right]}
\end{equation}
where C is the set of all connected cells, a and b are two connected cells, $a_x$ and $a_y$ are coordinates of cell a and $w_{ab}$ is the weight of the connection between a and b. In Traditional area-driven and time-driven placements, $w_{ab}$ is defined as the wirelength and a function of timing criticality of the net connecting cell a and b, respectively.

To have a variation-aware quadratic placement to reduce SER, defining $w_{ab}$ as a function of an SER metric and a variation map index may be meaningful. However, determining the WID variation value of a cell is not possible without having information about the cell locations. Therefore, WID variation map information cannot be directly used for defining cell weights in quadratic placements, because the WID variation value is a function of a cell location and cell locations are unknown until the placement algorithm would be finished. 

We address this challenge by proposing a heuristic method for imposing WID variation map information on the quadratic placement. So, $w_{ab}$ is defined as a function of an SER metric. This metric should be able to estimate the contribution of each cell to the total SER of the circuit. There are some SER metrics in the literature, but we choose PVW \cite{4}\cite{37}, as its results are quite close to Monte Carlo results and its computation time is very low.

Eq. \ref{eq:4} can be decomposed into $\phi = \phi_x + \phi_y$. So we narrow our analysis to the part that belongs to x coordinates. $\phi_x$ can be written in the following form:
\begin{equation} \label{eq:5}
\phi(x) = \frac{1}{2}\vec{x}^T Q_x \vec{x} + C_x \vec{x} + constant
\end{equation}
where $\vec{x}$ is the vector of x-coordinates of all cells, and Q and C are weighted connectivity matrix denoting connections between movable cells, and connection between movable cells and fixed cells, respectively. Constant value in Eq. \ref{eq:5} is due to the presence of fixed cells such as I/O pins.

Using the approach of GORDIAN algorithm \cite{38}, our algorithm includes successive phases of optimization and bipartitioning. After the $l$th phase of bipartitioning, center of gravity constraints of each sub-region is generated as:
\begin{equation} \label{eq:6}
A^{(l)} x = u^{(l)}
\end{equation}
where $u^{(l)}$ is a vector containing center of gravity coordinates of sub-regions and $A^{(l)}$ is a matrix constructed as:
\begin{equation} \label{eq:7}
\left\{
        \begin{array}{ll}
           \frac{area_u}{\sum_{u \in M_p}{area_u}} & \quad if\ u \in M_p\\
            0 & \quad Otherwise
        \end{array}
    \right.
\end{equation}
where $area_u$ is the area of cell $u$ and $M_p$ is a set of cells belong to the region $p$.

After construction of the constraints, the following quadratic optimization is formed. It is proved that this optimization is convex and has a global minimum \cite{38}.
\begin{equation} \label{eq:8}
min \left\{ \phi(x) = \frac{1}{2}\vec{x}^T Q_x \vec{x} + C_x \vec{x}\  |\  A^{(l)} x = u^{(l)} \right\}
\end{equation}

After solving the optimization and sorting the cells based on x-coordinates, bipartitioning is carried out and then new constraints are generated. In the next phases, the bipatitioning progress is repeated by alternating between sorting in horizontal (x-coordinates) and vertical (y-coordinates) directions. This consecutive QP optimization and bipartitioning is performed until each region contains an individual cell.

\subsection{Adding the Effects of WID Variations}
As mentioned earlier, WID variation information is not considered directly in the objective function of our QP placement. Rather, we present a sensitivity-based technique for efficient use of WID variations information which tries to avoid placing sensitive cells to soft errors, at some regions of the die, using WID variation map.

In this work, we use the VARIUS framework \cite{32} to apply WID variations to process parameters of circuits. In this framework it is supposed that the chip is partitioned into N×N rectangular segments with the same dimensions. In each segment, the process parameter is modeled as a random variable with normal distribution having zero mean and a standard deviation of $\sigma_{sys}$.
VARIUS models the systematic variations with a multivariate normal distribution and uses a spherical structure for modeling the spatial correlation. Its function of correlation is dependent on Euclidean distance between segments.

Channel length $(L_{eff})$ and threshold voltage $(V_{th})$ are two process parameters that are typically considered to be affected by WID variations. In the model of VARIUS \cite{32}, the relation between these two parameters is modelled as:
\begin{equation} \label{eq:3}
L_{eff} = L_{eff}^{n} \left(1 + \frac{V_{th} - V_{th}^{n}}{2V_{th}^{n}} \right)
\end{equation}
where $L_{eff}^{n}$ and $V_{th}^{n}$ are nominal values of $L_{eff}$ and $V_{th}$ respectively. As a result, we only consider the variations in $V_{th}$ in our work, as $L_{eff}$ can be calculated using Eq. \ref{eq:3}.

The extracted variation map of $V_{th}$ has N×N fragments where each fragment has a value of $V_{th}$ and $V_{th}$ values of cells placed at a fragment are equal to that of the fragment. Variations in $V_{th}$ values of cells lead to changes in their delays which can greatly affect the amount of electrical masking and subsequently the circuit SER.

As values of $V_{th}$ and $L_{eff}$ of gates are dependent on their locations due to the WID variations, calculating the gate propagation delay is not straightforward. We all know that the delay of a gate can be calculated as follows:
\begin{equation} \label{eq:9}
Delay_g = T_{INV} \left(LE_g + FO_g + P_g \right)
\end{equation}
where $T_{INV}$ is the intrinsic delay of an inverter, $LE_g$, $FO_g$ and $P_g$ are logical effort, electrical effort and parasitic delay of a gate $g$, respectively. The high dependency of $T_{INV}$ on $V_{th}$ can be explained by the well-known alpha-power law \cite{39}:
\begin{equation} \label{eq:10}
T_{INV} \propto \frac{L_{eff} V_{DD}}{(V_{DD} - V_{th})^{\alpha}}
\end{equation}
where $\alpha$ is a constant which is dependent on process technology. Substituting Eq. \ref{eq:3} and Eq. \ref{eq:10} into Eq. \ref{eq:12}, following equations can be derived:
\begin{equation} \label{eq:11}
T_{INV} \propto \frac{V_{DD} \left(1 + \frac{V_{th}}{V_{th}^{0}} \right)}{(V_{DD} - V_{th})^{\alpha}}
\end{equation}
\begin{equation} \label{eq:12}
Delay_g \propto \frac{V_{DD} \left(1 + \frac{V_{th}}{V_{th}^{0}} \right)}{(V_{DD} - V_{th})^{\alpha}} \left(LE_g + FO_g + P_g \right)
\end{equation}

The increase in delay of a gate leads the to increase of electrical masking of the gate, based on the electrical masking modeling of PVW \cite{4}\cite{37}. So, in order to decrease the circuit SER, cells should be placed with an approach that results in increase of circuit electrical masking. It can be empirically observed that Eq. \ref{eq:11} has a quite linear behavior with respect to  $V_{th}$ in the parameter range of interest. Therefore, placing cells in regions with high $V_{th}$ (HVT regions) reduce the circuit SER. To put it differently, for the sake of circuit SER reduction, placing cells in low $V_{th}$ regions (LVT regions) should be preferably avoided. 

However, this is not a feasible strategy for all cells because the die area is bounded and not all cells can be placed in HVT regions. Moreover, placing some cells in HVT regions should be carried out carefully, as it may negatively affect the critical path and the circuit delay. Therefore, we restrict this approach to sensitive cells to transient pulses.

Our aforementioned strategy includes three steps. First, sensitive cells to SER are identified. Second, chip fragments are classified with respect to the value of $V_{th}$. Finally, the QP equations applying the placement strategy are generated.

\subsubsection{Sensitive Cell Classification} \label{sec:g-cell-id}
Identifying sensitive cells to transient pulses needs an SER metric which has the ability to estimate the relative SER contribution of each individual cell to the circuit SER. In this regard, we choose the PVW model \cite{4}\cite{37} as an SER metric and use Eq. \ref{eq:13} for finding sensitive cells to soft errors.
\begin{equation} \label{eq:13}
SER_{A} \leq \frac{SER_{ckt}}{number\ of\ cells} \longrightarrow A\ is\ sensitive.
\end{equation}
where $SER_{ckt}$ is the total circuit SER and $SER_A$ is the individual SER of cell A.
\subsubsection{Region Classification}
Performing our placement strategy needs two sets of chip fragments classified with respect to the $V_{th}$ value extracted from the variation map. Table \ref{tab:1} presents the empirically obtained criteria of chip regions classification in which $\mu_{V_{th}}$ and $\sigma_{V_{th}}$ are the mean and the standard deviation of $V_{th}$ used for constructing variation map, respectively.

\begin{table}[h!b!t!p!]
\centering
\vspace{-0.05in}
\caption{Definition of HVT and LVT Regions}
\vspace{-0.05in}\label{tab:1}
\centering
\begin{tabular}{|l|c|c|c|} 
\cline{2-4}
\multicolumn{1}{l|}{} & Nominal Value & HVT region & LVT region  \\ 
\hline
        $V_{th}$              & 220 mv & $\mu_{V_{th}} + 1.3\sigma_{V_{th}} < V_{th}$ &  $\mu_{V_{th}} + 1.3\sigma_{V_{th}} \geq V_{th}$ \\
\hline
\end{tabular}
\end{table}

\begin{figure}[!tb]
	\begin{center}
		\includegraphics[width = 0.47\textwidth]{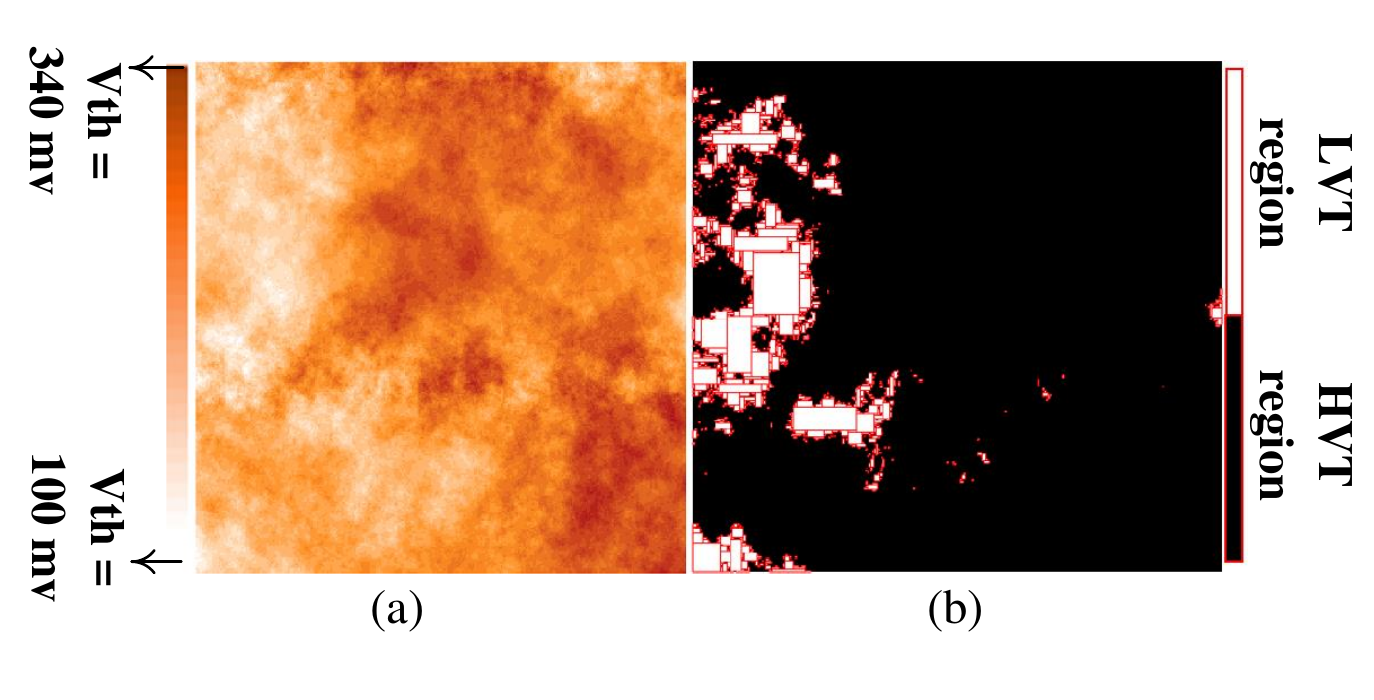}
		\vspace{-0.2in}
		\caption{a) $V_{th}$ variation map of a typical die and, b) its corresponding R-map.}
		\vspace{-0.2in}
	\label{fig:3}
	\end{center}	
\end{figure}

By applying the definition presented in Table \ref{tab:1} to the variation map of the chip, we can generate a map named as LH-map. Due to the spatial correlation, there are many regions in the map in which adjacent fragments are in the same class. To simplify the formulating of our placement strategy used for considering WID variations, adjacent fragments belonging to the same class are merged leading to creation of some LVT and HVT blocks. We deploy rectangular map (R-map) \cite{40} to address this issue.
We consider LH-map as a binary matrix in which LVT and HVT fragments are denoted by 0 and 1, respectively, to obtain R-map representations. Giving this binary matrix as an input to R-map, it traverses the matrix and merges adjacent 0 and 1 to create maximal blocks of 0 and 1 \cite{40}.The constructed variation map and its equivalent R-map is shown in  Figure \ref{fig:3}.

\subsubsection{Optimization Formulation}
Forming the optimization can be performed after classifying regions and cells. As mentioned previously, the algorithm should apply the placement strategy in which placing sensitive cells in LVT regions is preferably avoided. To this end, a penalty-based approach is used in which placing sensitive cells in LVT regions results in adding penalties to the objective function. Actually, using this approach the algorithm tries to make trade-off between decreasing the circuit SER by not placing sensitive cells in LVT regions and reducing the total wirelength by placing sensitive cells in nearby LVT regions.

To formulate the proposed placement strategy, let V and B be the sets of all cells and all LVT blocks, respectively. We assume that the set V is partitioned into two subsets $V_S$ and $V_{NS}$, respectively including sensitive cells and other cells. Eq. \ref{eq:14} presents the proposed objective function in which the second sigma sign is for calculation of imposed penalties caused by placing sensitive cells in LVT blocks.
\begin{equation} \label{eq:14}
\begin{split}
\phi = \frac{1}{2} \sum_{(m,n) \in V} (1 - {SER_{m}^{'}}) ( (x_m - x_n)^2 + (y_m - y_n)^2) \\ + \sum_{i \in V_{S}\ and\ j \in B}{\delta_{ij} \times e^{{(SER_{i}^{'}} + 1)^K}}
\end{split}
\end{equation}
where ${SER_{m}^{'}}$ is the normalized SER of cell m (SER of all circuit cells are normalized between 0 to 1), $\delta_{ij}$ parameter is for identifying whether cell $v_i$ is placed at LVT block $b_j$ or not, and K is a constant parameter used for controlling the amount of penalties which is obtained after extensive experiments.
\begin{equation} \label{eq:15}
\delta_{ij} = \left\{
        \begin{array}{ll}
            1 & \quad if\ v_i\ is\ placed\ in\ b_j\\
            0 & \quad if\ v_i\ is\ not\ placed\ in\ b_j
        \end{array}
    \right.
\end{equation}

Eq. \ref{eq:16} defines an LVT block $b_j$ in which $l_{x_j}$ and $u_{x_j}$ are its lower and upper bounds in x-axis and $l_{y_j}$ and $u_{y_j}$are the same for y-axis.
\begin{equation} \label{eq:16}
b_j = \left\{ (x,y)\ |\ l_{x_j} \leq x \leq u_{x_j}\ and\ l_{y_j} \leq y \leq u_{y_j} \right\}
\end{equation}

Supposing $v_i$ as a cell with $(x_i,y_i)$ coordinates, for modeling the proposed placement strategy the following condition should be satisfied: $\delta_{ij}=1$, if and only if the cell $v_i$ is placed at block $b_j$; otherwise, $\delta_{ij}=0$. 

We introduce Eq. \ref{eq:17} in order to present constraints for modeling the aforementioned condition for $x_i$ in which $\delta_{ij}^1,\delta_{ij}^2$ and $\delta_{ij}^3$ are temporary binary variables and M has a value greater than the maximum value of x and y coordinates of the chip. Generated constraints of $y_i$ are the same.
\begin{equation} \label{eq:17}
\begin{split}
&-M(1 - \delta_{ij}^1) + l_{x_j} \leq x_i < l_{x_j} + M \delta_{ij}^1\\
&-M\delta_{ij}^2+ u_{x_j} < x_i \leq u_{x_j} + M(1 - \delta_{ij}^2)\\
&0 \leq \delta_{ij}^1 + \delta_{ij}^2 - 2\delta_{ij}^3 \leq 1\\
&1 - M(1 - \delta_{ij}^3) \leq \delta_{ij} \leq 1 + M(1 - \delta_{ij}^3)\\
&-M\delta_{ij}^3 \leq \delta_{ij} \leq M\delta_{ij}^3\\
&\forall i \in V_S,\ \forall j \in B 
\end{split}
\end{equation}

These constraints are generated for all LVT blocks, however some of these LVT blocks may be single and scattered as shown in Figure \ref{fig:3}. It was observed that generating aforementioned constraints for these scattered blocks increases the algorithm complexity and the total wirelength overhead of the placement in most cases. Therefore, prior to generating constraints, some single and scattered LVT blocks are omitted and a few of HVT blocks, which partitioned big LVT blocks into small blocks, are counted as LVT blocks. Putting all these together, the pseudo code of our SeaPlace-G algorithm is shown in Algorithm 1.

\begin{figure}[!t]
	\begin{center}
		\includegraphics[width = 0.48\textwidth]{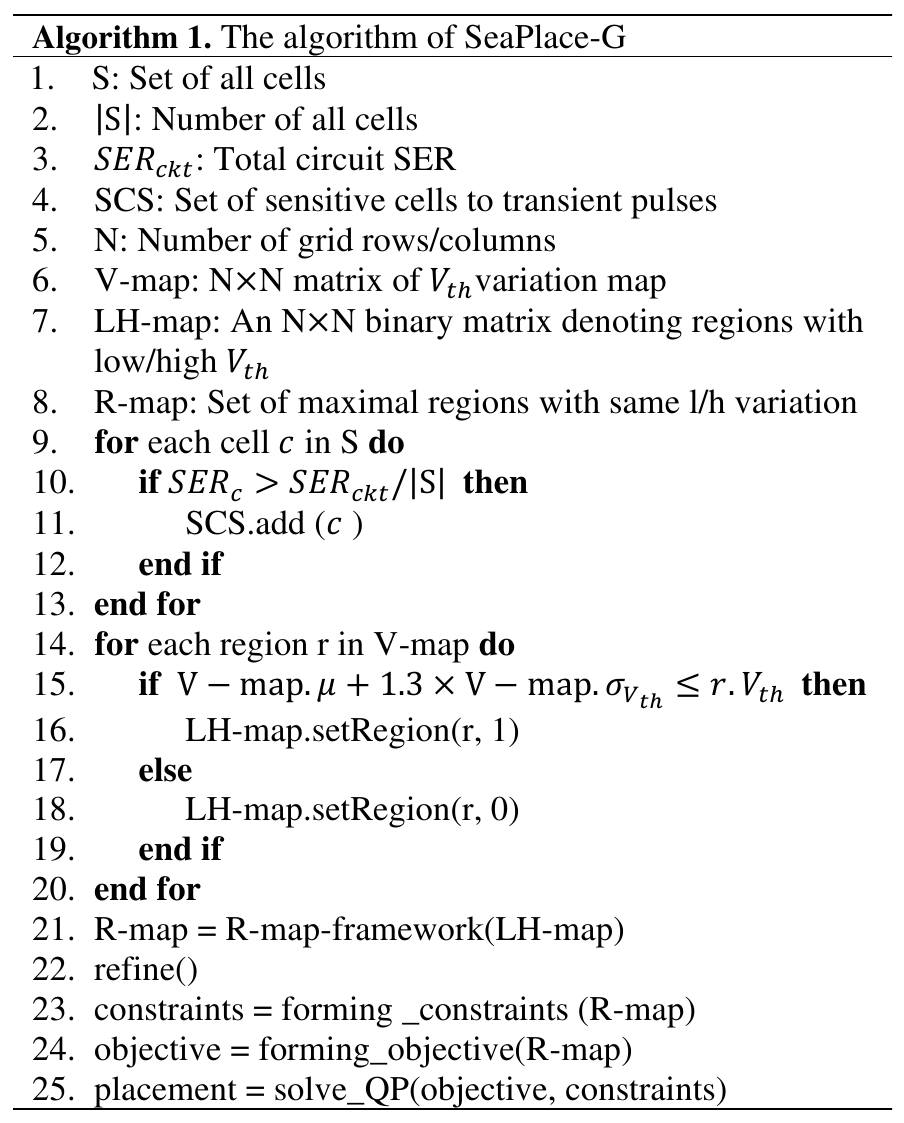}
		\vspace{-0.2in}
	\end{center}	
\end{figure}
\section{SeaPlace-D: A Variation-Aware and MET-Aware Detailed Placement Algorithm} \label{medp}

\begin{figure}[!tb]
	\begin{center}
		\vspace{-0.15in}		\includegraphics[width = 0.3\textwidth]{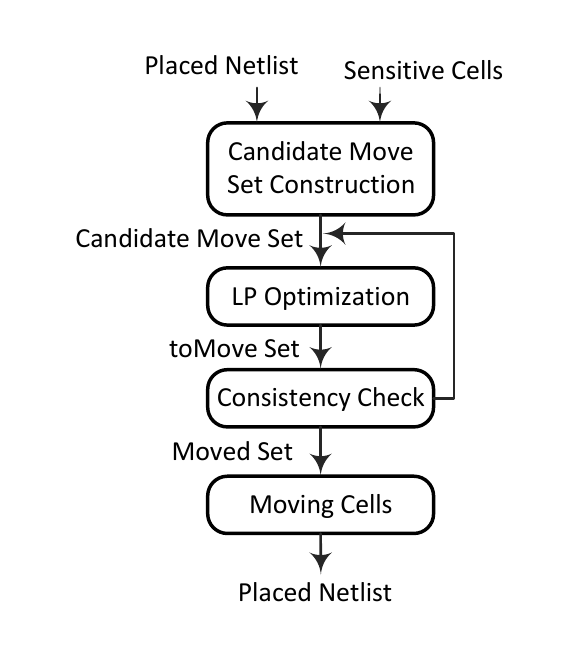}
		\vspace{-0.25in}
		\caption{Overall flow of the SeaPlace-D algorithm.}
		\vspace{-0.2in}
	\label{fig:SeaPlace}
	\end{center}	
\end{figure}

In this section we introduce a detailed placement that is hardened against METs and shown in Figure \ref{fig:SeaPlace}. This algorithm receives sensitive cells and placed netlist generated by SeaPlace-G placement. In the first phase of the algorithm, a set of pair candidates for moving are constructed. This set is named as \emph{Candidate Move SET (CMS)}. Then, an linear programming (LP) optimization is applied to CMS and build the \emph{toMove set}, a set of moves with high potential of increasing hardening and satisfying overhead constraints. The next phase, consistency check, evaluates the toMove Set and selects feasible moves as the \emph{moved set}. The moved sets are moved, and other moves that are not accepted by consistency check are served to LP optimization as input. This process will execute until some terminating conditions are met.

\begin{figure}[t]
	\begin{center}
		\includegraphics[width = 0.25\textwidth]{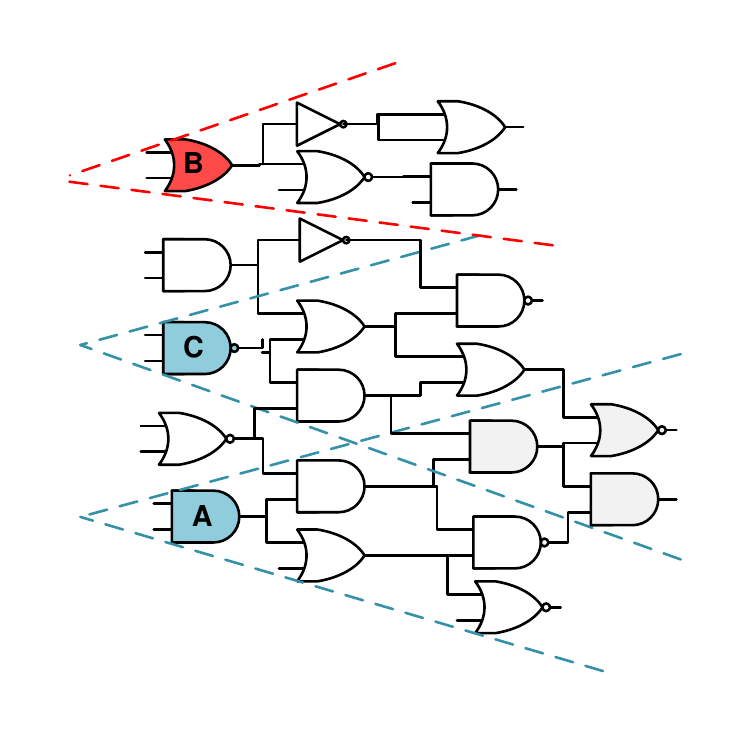}
		\vspace{-0.2in}
		\caption{Example of dependent and independent failure probability: A and B are independent, A and C are dependent.}
		\vspace{-0.2in}
	\label{fig:4}
	\end{center}	
\end{figure}

\subsection{Multiple Event Transient Model}
Generally, our placement identifies some specific pairs of cells and tries to decrease the distance between cells of each pair. In this work, we only consider double event transients (DETs) because the probability of occurring more than two transient faults in the target technology of this work (45 nm) is about five percent \cite{24}.

As it was shown in  \cite{24},  if a particle strike affects two adjacent cells, their propagated errors may mask each other. The reason is that transient pulses generated by the strike may reach to a gate and cancel out each other. 
Figure \ref{fig:4} shows an example. In this figure cells A and B are independent, i.e., they share no gates in their forward cones. As a result, if a particle strike affects these two cells, their errors can propagate independently. So the total failure probability that a strike affects A and B is:
\begin{equation} \label{eq:18}
FP_{A,B}^{ind}  = FP_A+ FP_B - FP_{A \cap B}= FP_A + FP_B - FP_A . FP_B
\end{equation}
where $FP_A$ and $FP_B$ are the failure probabilities caused by affecting A and B. As $FP_A$ and $FP_B$ are independent, their intersection is equal to their product. We denote this probability with $FP_{A,B}^{ind}$ in the rest of this paper and refer to it as the joint independent failure probability. 

As shown in Figure \ref{fig:4}, cells A and C are not independent, as they share some gates in their forward cones. We denote the failure probability of two dependent cells like A and C with $FP_{A,C}$ and name it as the joint failure probability (JFP) which is calculated using the method in \cite{41}. As the propagated errors from two affected cells may meet each other, their JFP may be changed. If two transient pulses, originated from two error sites, reach to a gate, their interaction can result in weaker or stronger pulses. So, depending on the circuit structure, having intersection between forward cones of error sites may or may not cause lower JFP. It is shown that the occurrence of this pulse shrinking, known as pulse quenching effect, is closely relevant to electrical relationship between physically adjacent cells and their distances \cite{25}. Hence, identifying electrically related cells that can have such error propagation behaviors and making them physically closer, can be advantageous to mitigate MET SER. We call these cells as \emph{candidate pairs (CAP)} and a set of all CAPs as \emph{CPS}.

Reducing the distance between cells of each CAP to use their possible quenching effects should be done with respect to the patterns of MET occurrence. In this regard, the cells should be located in a distance in which hitting a particle to one of them can affect the other one. Similar to \cite{24}, we use an oval shape as the shape of the affected area for MET occurrence. Hence, to have a CAP, the two cells should be at a distance where cells can be fitted in the oval. The maximum amount of this distance is called as \emph{masking distance}. In this paper, we only consider the horizontal distance as a masking distance, as the existence of VDD or GND trail between two consecutive rows leads to vertical distances bigger than the masking distance according to our experiments. In the following we explain our placement strategy.

\subsection{Candidate Pair Identification}
To identify CAPs we use an approach inspired by \cite{24}. As mentioned earlier, when a particle strike happens, we are interested in scenarios that propagated errors from affected cells, mask each other and result in failure probability reduction. In other words, in order to find a CAP, we are looking for two cells that their JFP is lower than their joint independent failure probability. So, the joint independent failure probability of all possible pairs should be calculated in the first step. Calculation of JFP for all possible pairs of a circuit has a time complexity of order $O(n^2)$ that is very time consuming for large circuits. In order to facilitate this calculation, we narrowed the search for CAPs to sensitive cells to transient pulses. More precisely, each sensitive cell $c$ and the cells connected to its input and output pins are used to build our new search space. The reason of considering cells connected to each sensitive cell is that the expected pulse quenching is more probable especially for cells that have fan-in and fan-out relationships \cite{25,42}. After calculating JFP for each connected pair (A-B) of our search space, we compare $FP_{(A-B)}$ and $FP_{(A-B)}^{ind}$. If the former is less than the latter, A and B form a CAP, i.e.,:
\begin{equation} \label{eq:19}
FP_{(A-B)} < FP_{(A-B)}^{ind}  \iff  (A-B) \in CPS
\end{equation}

\subsection{Sensitive Cell Identification}
To identify sensitive cells to transient pulses, we apply the Eq. \ref{eq:13}, which is the same process as discussed in Section~\ref{sec:g-cell-id}.

\subsection{Sensitive Cell Transfer and Swap}
After identifying CAPs, we intend to make them physically closer in the layout to use their masking effects. Our algorithm uses two types of actions to reduce the distance between cells of each CAP that named as \emph{transfer} and \emph{swap}. Before going into the details, we propose a $\Omega$ notation for denoting different types of actions. The superscript of $\Omega$ operator is used to write cells of a pair with their final coordinates after doing the action and the subscript is for mentioning the cells supposed to move with their coordinates before doing the action.

\begin{figure}[!tb]
	\begin{center}
		\includegraphics[width = 0.45\textwidth]{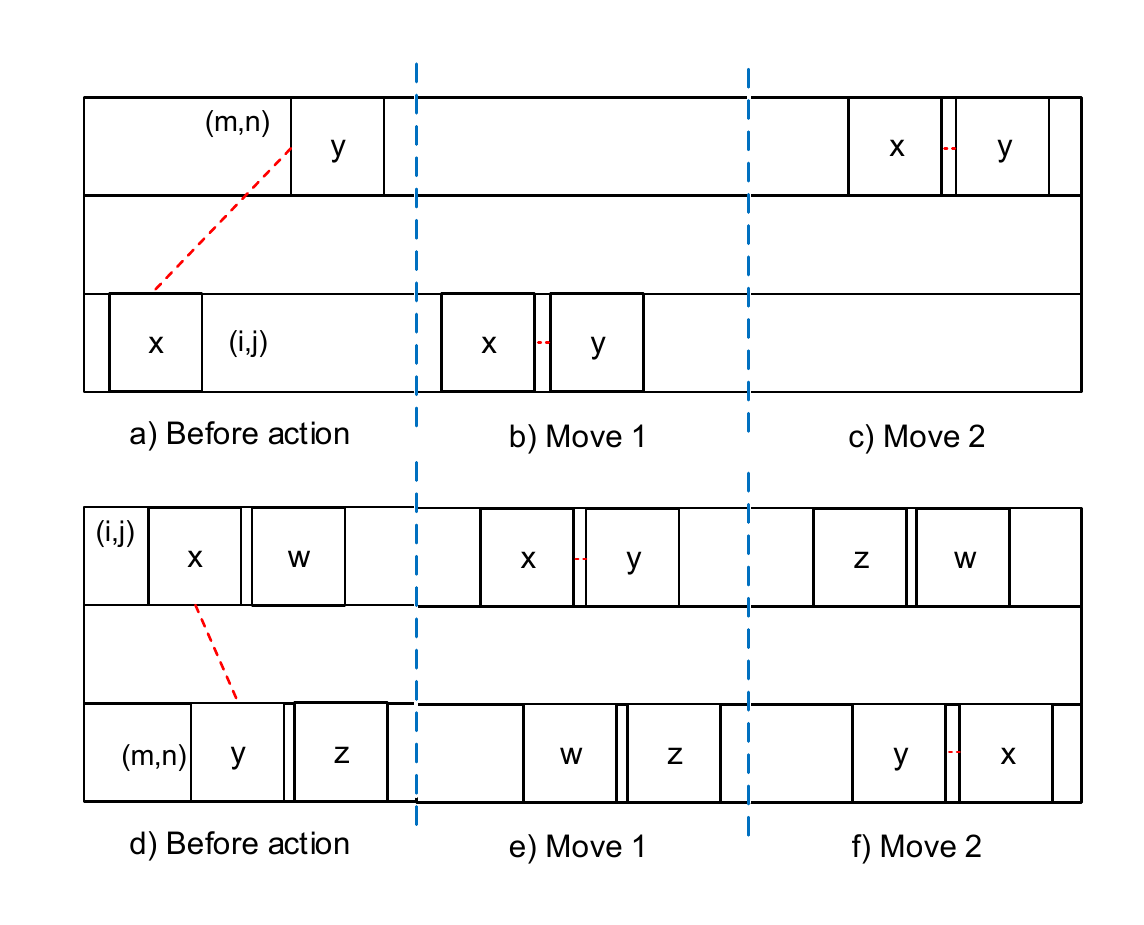}
		\vspace{-0.1in}
		\caption{Examples of transfer (a-c) and swap (d-f) actions. Note that move 1 and 2 in this figure (and all following figures) are two different move options applied to the original cells; they are not applied on top of each other.}
		\vspace{-0.2in}
	\label{fig:5}
	\end{center}	
\end{figure}

The transfer and swap actions place the cells of each CAP at a distance less than the masking distance. Figure \ref{fig:5}a shows two cells of a CAP placed at separate rows. Reducing the distance between these two cells can be done in two different moves as shown in the Figure \ref{fig:5}b and Figure \ref{fig:5}c. Moves 1 and 2 (called transfers in this case) are formulated by Eq. \ref{eq:20} and Eq. \ref{eq:21}, respectively:
\begin{equation} \label{eq:20}
\Omega_{c_y^{m,n}-c_V^{i,j+1}}^{c_x^{i,j}-c_y^{i,j+1}} \left(c_y^{m,n}, c_V^{i,j+1}, c_x^{i,j} \right)= \left( c_x^{i,j} - c_y^{i,j+1}, c_V^{m,n} \right)
\end{equation}
\begin{equation} \label{eq:21}
\Omega_{c_V^{m,n-1} - c_x^{i,j}}^{c_x^{m,n-1} - c_y^{m,n}} \left(c_V^{m,n-1}, c_x^{i,j}, c_y^{m,n} \right) = \left( c_x^{m,n-1}, c_y^{m,n}, c_V^{i,j} \right)
\end{equation}
where $c_x^{i,j}$ denotes cell $x$ that is placed at $(i,j)$ coordinates of the layout grid. As only one cell is moved in a transfer action, the vacant position in $(i,j+1)$ coordinates in Figure \ref{fig:5}a, which is supposed to be filled by cell $c_y^{m,n}$, is denoted by $c_V^{i,j+1}$ in Eq. \ref{eq:20}.

Swap action is performed in circumstances that it is not possible to place the cells close to each other by using a transfer action. An example is shown in Figure \ref{fig:5}d. In this example, placing cells x and y at a very close distance is not possible due to the lack of sufficient space around both cells. In this situation $\frac{x}{y}$ swaps its location with that of the neighbor of $\frac{y}{x}$. These two different moves (called swaps in this case) are shown in Figure \ref{fig:5}e and Figure \ref{fig:5}f and are described by Eq. \ref{eq:22} and Eq. \ref{eq:23} respectively:
\begin{equation} \label{eq:22}
\Omega_{c_y^{m,n} - c_w^{i,j+1}}^{c_x^{i,j} - c_y^{i,j+1}} \left( c_y^{m,n}, c_w^{i,j+1}, c_x^{i,j} \right) = \left( c_x^{i,j}, c_y^{i,j+1}, c_w^{m,n} \right)
\end{equation}
\begin{equation} \label{eq:23}
\Omega_{c_x^{i,j} - c_z^{m,n+1}}^{c_y^{m,n} - c_x^{m,n+1}} \left( c_x^{i,j}, c_z^{m,n+1}, c_y^{m,n} \right) = \left( c_y^{m,n}, c_x^{m,n+1} ,c_z^{i,j} \right)
\end{equation}

\begin{figure*}[!tb]
	\begin{center}
		\includegraphics[width =0.7\textwidth]{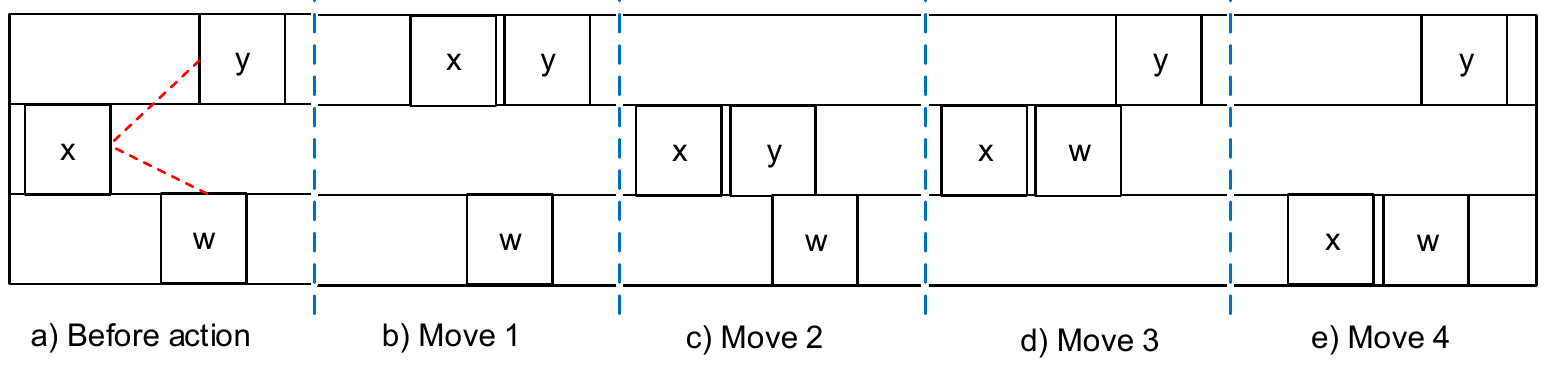}
		\vspace{-0.1in}
		\caption{An example of multiple move alternatives.}
		\vspace{-0.2in}
	\label{fig:6}
	\end{center}	
\end{figure*}

As shown in Figure \ref{fig:5}, making the cells of a pair closer can be done in different ways. There can be more ways if CAPs share some cells. In Figure \ref{fig:6}, cell x is common between $(x,y)$ and $(x,w)$ pairs. In this example, there are four moves that can make cells of a CAP closer. Thus, each transfer or swap action can be performed through different moves and it is up to the algorithm to select one of these moves. A set that contains all possible moves for making cells of each CAP closer is named as \emph{Candidate Move Set (CMS)}.

\begin{figure}[!tb]
	\begin{center}
			\vspace{-0.05in}	\includegraphics[width = 0.42\textwidth]{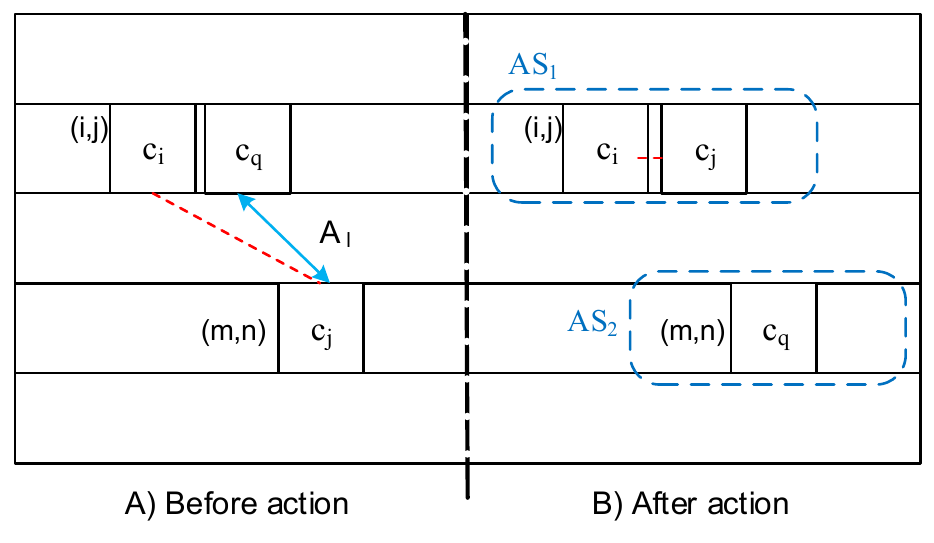}
		\vspace{-0.15in}
		\caption{Two ASes resulting from the swap action $A_l$.}
		\vspace{-0.1in}
	\label{fig:7}
	\end{center}	
\end{figure}

After forming CMS, we calculate the delta failure probability $(\Delta FP)$ for all of CMS members. Delta failure probability is a parameter that approximates the amount of changes in the failure probability of a circuit after performing a transfer or a swap action. In order to calculate delta failure probability, we first define \emph{Affected Sets (AS)}. When a change is made to a point of the layout by an action, a pair of cells where their JFP has changed after the action, form a pair that belongs to AS. After performing a transfer or a swap action, two ASes are formed. The ASes of a swap action are depicted in Figure \ref{fig:7}. For calculating the delta failure probability, the changes made to the JFP of all members of each AS is calculated. Delta failure probability of Figure \ref{fig:7} is calculated as
\begin{equation} \label{eq:24}
\begin{split}
&\Delta FP{A_l^{c_i - c_j }}= \sum_{(r,s) \in (AS_1 \cap AS_2)} FP_{(r-s)}^2 - FP_{(r-s)}^1\\
&= \sum_{(r,s) \in AS_1} FP_{(r-s)}^2 - FP_{(r-s)}^1 + \sum_{(r,s) \in AS_2} FP_{(r-s)}^2 - FP_{(r-s)}^1\\
&= \sum_{r = (c_i\ or\ c_j)} \sum_{s \in AS_1} FP_{(r-s)}^2 + \sum_{t \in AS_2} FP_{(c_q - t)}^2\\
&- \sum_{r = (c_i\ or\ c_q)} \sum_{s \in AS_1} FP_{(r-s)}^1 - \sum_{u \in AS_2} FP_{(c_j - u)}^1
\end{split}
\end{equation}
where (r-s) is a pair, $A_l^{c_i - c_j}$ is the $l$th move of an action related to cells $c_i$ and $c_j$ shown in Figure \ref{fig:7}, and $FP_{(r,s)}^1$ and $FP_{(r,s)}^2$ are the joint failure probabilities of cells of pair (r,s), before and after an action.

Computation of delta failure probability in Eq. \ref{eq:24} does not need the calculation of all joint failure probabilities. It can be executed in a quick time, by using the recently calculated values. By saving JFP values of cells that are calculated in CAPs identification phase, required JFP values in Eq. \ref{eq:24} can be obtained quickly.

Changing the location of cells in the layout may have a negative effect on the SET SER of the circuit, especially when the location of sensitive cells are changed. So, in order to escape from a vicious circle caused by jeopardizing the achievements of global placement, we need a mechanism to maintain the achievements of our global placement in hardening against SETs. We do this by measuring the changes made to the SER of the circuit after performing each action while putting a constraint on the total allowed change of circuit SER in our optimization algorithm. In Figure \ref{fig:7}, $\Delta SER$, that is the difference between the SER of the circuit before and after doing an action, is computed as follows:
\begin{equation} \label{eq:25}
Delta SER \left( A_l^{c_i-c_j} \right) = \Delta SER (c_j) + \Delta SER (c_q)
\end{equation}
where $\Delta SER (c_j)$ is the change made to SER of $c_j$ due to the change of its location. Wirelength is another constraint that we consider in our algorithm aiming to control the imposed overhead by actions.
\begin{equation} \label{eq:26}
\Delta WL \left( A_l^{c_i-c_j} \right) = \Delta WL (c_j) + \Delta WL(c_q)
\end{equation}
where $\Delta WL (c_j)$ is the change made to wirelength of $c_j$ due to the change of its location. Algorithm 2 shows the pseudo-code description of our algorithm for construction of CAPS and CMS. The algorithm has a linear time complexity of $O(n)$.

\begin{figure}[!t]
	\begin{center}
		\includegraphics[width = 0.48\textwidth]{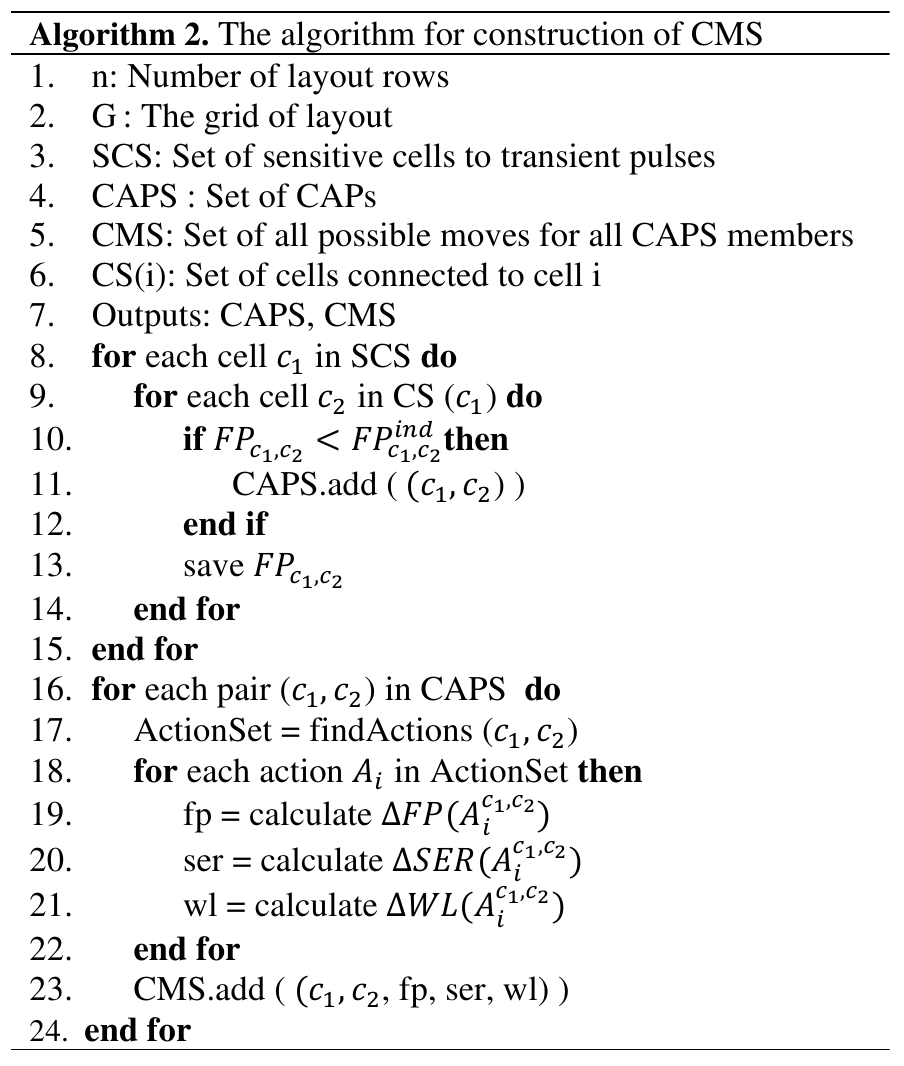}
		\vspace{-0.2in}
	\end{center}	
\end{figure}

\subsection{Placement Algorithm}
After calculating overhead and reliability related parameters for each move, we can apply an algorithm for performing the moves. During our experiments, we observed that performing the moves with negative and high delta failure probability increases the chance of pulse quenching effect and results in a considerable total failure probability reduction for the circuit. To this end, we perform an optimization over CMS. We propose an LP optimization that maximize the sum of delta failure probability of moves for the solution, while imposing some constraints to the delta SER and delta wirelength of the circuit. $SER_{total}$ and $WL_{total}$ are the total amounts of SER and wirelength of the circuit obtained before performing the optimization and $\Delta SER$ and $\Delta WL$ are the maximum allowed percentage of change in SER and wirelength respectively. To formulate the optimization, we suppose that CAPS has $n$ elements and each element (each CAP) can be performed by at most \emph{m} moves. In the following equations, $\delta_{ij}$ is a temporary binary parameter used to specify actions which are present in the solution set.
\begin{equation} \label{eq:27}
min \left\{ - \sum_{i=1\ to\ n,\ j=1\ to\ m} \delta_{ij} \times \Delta FP \left( A_j^{Pair(c_p, c_q)} \right) \right\}
\end{equation}
Subject to:
\begin{equation} \label{eq:271}
\begin{split}
&\forall i, 1 \leq i \leq n, \sum_{j=1}^m \delta_{ij} = 1\\
&\forall i, \forall j, 1 \leq i \leq n, 1 \leq j \leq m,\\
&\delta_{ij} \times \Delta SER \left( A_j^{Pair (c_p, c_q)} \right) \leq (1 + \Delta SER) \times SER_{total}\\
&\delta_{ij} \times \Delta WL \left( A_j^{Pair(c_p,c_q)} \right) \leq (1 + \Delta WL) \times WL_{total}
\end{split}
\end{equation}

\begin{figure}[!tb]
	\begin{center}
		\includegraphics[width = 0.45\textwidth]{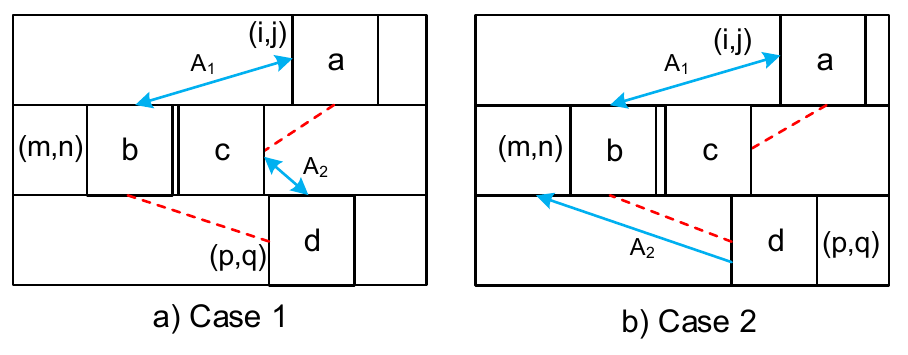}
		\vspace{-0.15in}
		\caption{Examples of inconsistency between moves}
		\vspace{-0.2in}
	\label{fig:8}
	\end{center}	
\end{figure}

By solving the optimization problem in Eq. \ref{eq:27}, a solution set of moves is obtained. However, we may not be interested to perform all moves of the solution set. This is because of an inconsistency between elements of the solution set. Indeed, doing a move is not independent and it depends on previously performed moves of the solution set. So, performing a move is beneficial, if suitable circumstances exist. Two different examples of inconsistency between the solution moves are depicted in Figures \ref{fig:8}a and \ref{fig:8}b. Both examples (Case 1 and Case 2) include two actions A1 an A2 and show a swap-swap and a swap-transfer actions respectively. We suppose that the move of action A1 is already performed and we are going to perform the action A2. But performing A2 is not effective, because in both examples the fix cell (cell b) of A2 related pair (pair (b-d)) has been moved to another location by A1. So, performing of A2 do not make cells of pair (b-d) physically closer. As a result, before doing a move of the solution set, the consistency between the move and already performed moves should be checked. We figure out this consistency checking challenge by presenting a consistency condition, called as Triple Condition. This condition should be satisfied in $\Omega$-based calculations.

\textbf{Triple Condition:} If two cells of subscripts and one of cells in superscripts of an $\Omega$ operator exist in its operand set, the triple condition is met and this move is consistent with previously performed moves.
\begin{equation} \label{eq:28}
\begin{split}
(c \in M \land d \in M) &\land (a \in M \lor b \in M))\\
&\longrightarrow \Omega_{c-d}^{a-b}(M) is consistent
\end{split}
\end{equation}

Therefore, consistency checking of moves of solution set can be done by checking the triple condition for each move. For example, the consistency verifying of Figure \ref{fig:8}a is as follows:
\begin{equation} \label{eq:29}
\begin{split}
& A_1 = \Omega_{{c_b}^{m,n} - c_a^{i,j}}^{c_a^{m,,n} - c_c^{m,n+1}} \left( c_b^{m,n}, c_a^{i,j}, c_c^{m,n+1} \right) = \left( c_a^{m,n}, c_c^{m,n+1}, c_b^{i,j} \right)\\
& A_2 = \Omega_{c_d^{p,q} - c_c^{m,n+1}}^{c_b^{m,n} - c_d^{m,n+1}} \left(c_d^{p,q}, c_c^{m,n+1}, c_b^{i,j} \right) = \phi
\end{split}
\end{equation}

As shown in Eq. \ref{eq:29}, action $A_2$ cannot be performed, because none of the cells in superscript of $\Omega$ exist in the operand set of $\Omega$. After checking triple condition for all moves of the solution set, some moves may not be consistent with performed moves. So, a part of the solution set is not usable and it will be worse when a considerable part of the solution set is inconsistent with performed moves. To resolve this challenge, we sort the moves of optimization solution based on the delta failure probability. We then perform the moves that satisfied the triple condition starting from the move with maximum amount of delta failure probability. Cells that are not consistent with already performed actions are saved and are treated as a new CAPS in the next iteration of algorithm.

\begin{figure}[!t]
	\begin{center}
		\includegraphics[width = 0.48\textwidth]{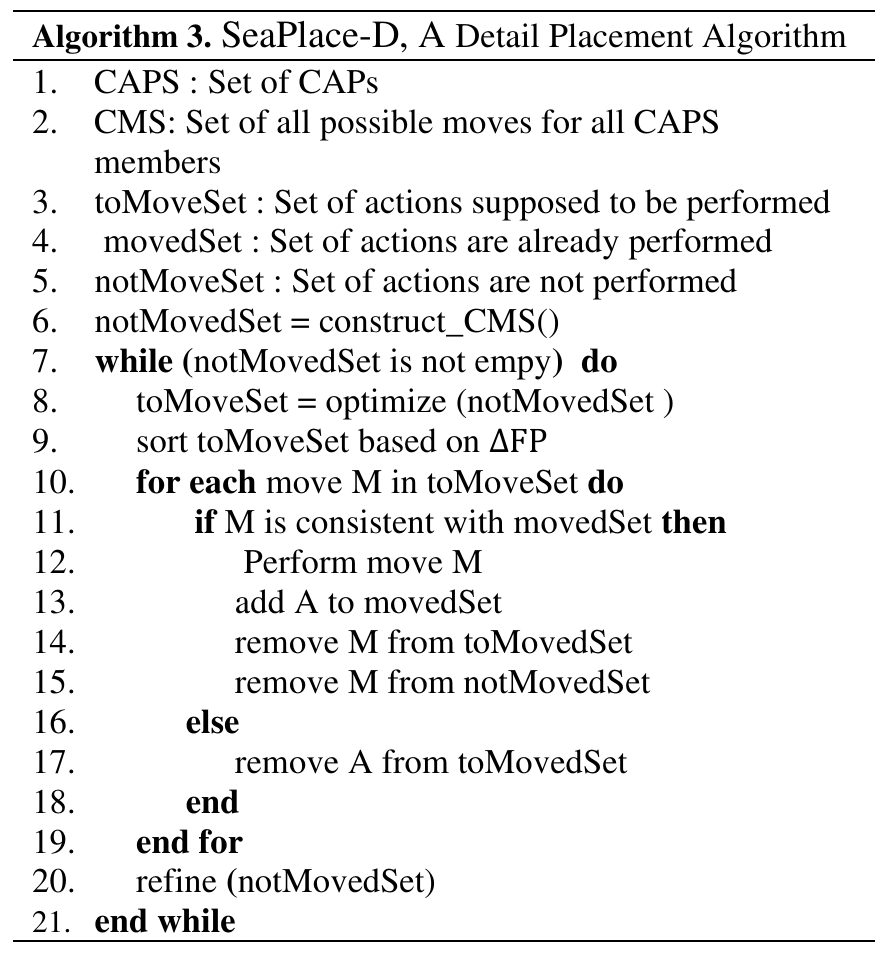}
		\vspace{-0.2in}
	\end{center}	
\end{figure}

The pseudo-code of our proposed detail placement algorithm is presented in Algorithm 3. The notMoved set is initialized by CAPS in the beginning of the algorithm and then inconsistent cells are added to it during the main loop of algorithm. As mentioned earlier, the optimization algorithm may be executed in several iterations, because of inconsistent moves. At the end of each iteration, the max allowed overheads are updated for the next iteration. The refine method in line 20 is responsible for terminating the algorithm execution by modifying notMovedSet in cases such as reaching the max allowed overhead and avoiding non-convergence.

\section{Experimental Results} \label{results}
\subsection{Setup}
The proposed algorithms are implemented in Java and applied to EPFL \cite{43} benchmark circuits for 45-nm Nangate technology. All simulations have been run on a Microsoft Windows machine with a Pentium Core i7 (2.1-GHz) processor and a 8-GB RAM. 

For constructing a variation map of $V_{th}$, we assume that the chip is partitioned into a N×N grid and run VARIUS with $\mu = 0.22$ \cite{33}, $3\sigma = 55\%$ \cite{33}, N = 300 and correlation distance $\phi=0.5$. The result is a variation map of systematic component of $V_{th}$. In this work, it is supposed that systematic and random components of $\sigma (V_{th})$ are equal; that is $\sigma_{sys} = \sigma_{rand} = \frac{\sigma}{\sqrt{2}}$ \cite{34}. We then generate a N×N matrix of samples from a zero-mean normal distribution with $\sigma_{rand} (V_{th})$. The final $V_{th}$ map is constructed by superposing systematic and random maps segment by segment.

The controlling parameters of Plcae-G and SeaPlace-D (Eq.~\ref{eq:271} and Eq.~\ref{eq:14}) were set as: $K$= 2.5 , $\Delta SER$= 0.1  and $\Delta WL$= 0.1.
To estimate SET-originated SER the method presented in \cite{37} is used as its accuracy is close to the Monte Carlo results and has a low runtime. We use the method introduced in \cite{24} for evaluating the achieved hardening against METs. 

\begin{figure}[!tb]
	\begin{center}
		\includegraphics[width = 0.48\textwidth]{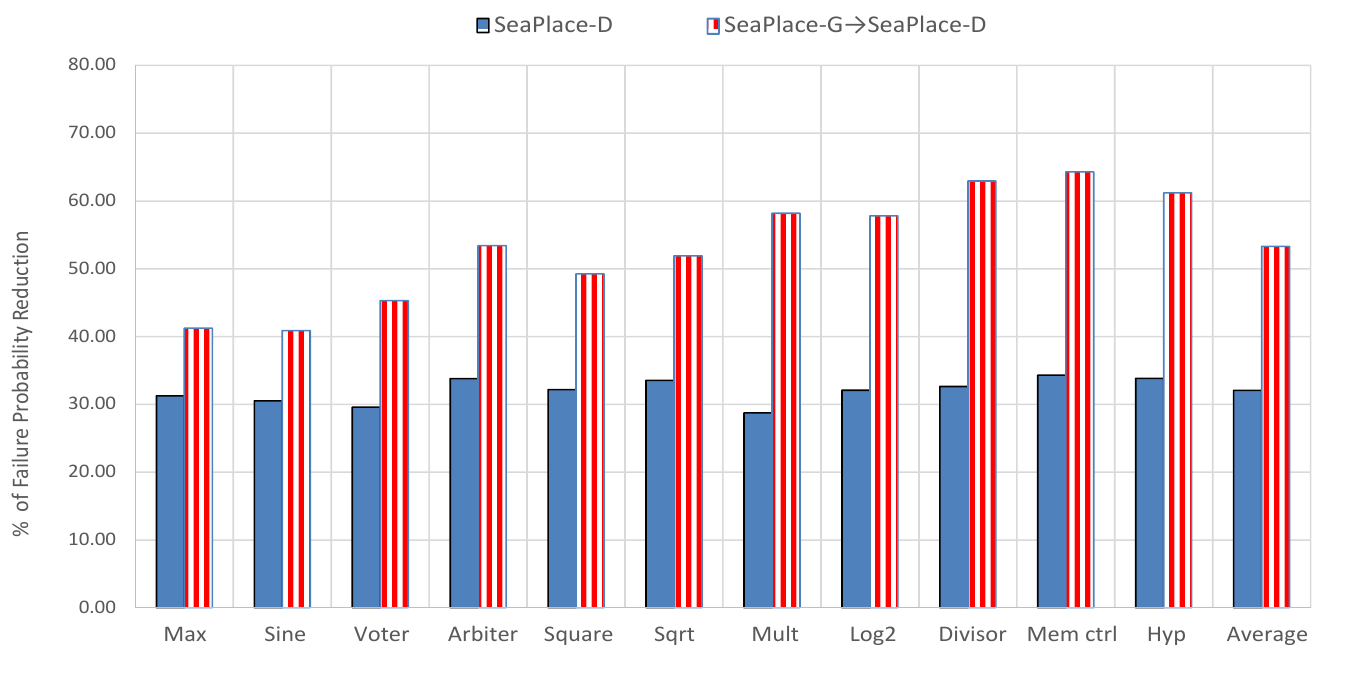}
		\vspace{-0.1in}
		\caption{Percent of failure probability reduction for SeaPlace-D and SeaPlace-G$\rightarrow$SeaPlace-D.}
		\vspace{-0.2in}
	\label{fig:9}
	\end{center}	
\end{figure}

\subsection{SER Reduction and Failure Probability Mitigation}
In this section, we evaluate the amount of failure probability mitigation achieved by our algorithms. Figure \ref{fig:9} shows the failure probability reduction for two cases. In the case denoted by SeaPlace-G→SeaPlace-D, SeaPlace-G and subsequently SeaPlace-D are applied to circuits while the other case reports the results of applying only SeaPlace-D. The results show that applying SeaPlace-D after an SeaPlace-G averagely results in about 53\% reduction in the failure probability while using only SeaPlace-D achieves about 32\% failure probability reduction, on average. This experiment indicates the efficiency of applying SeaPlace-G prior to SeaPlace-D on hardening against METs. Actually, performing SeaPlace-G not only reduces the amount of SET-originated SER, but also results in more significant failure probability reduction achieved by subsequent SeaPlace-D because of increasing the masking effects.

\begin{figure}[!tb]
	\begin{center}
		\includegraphics[width = 0.48\textwidth]{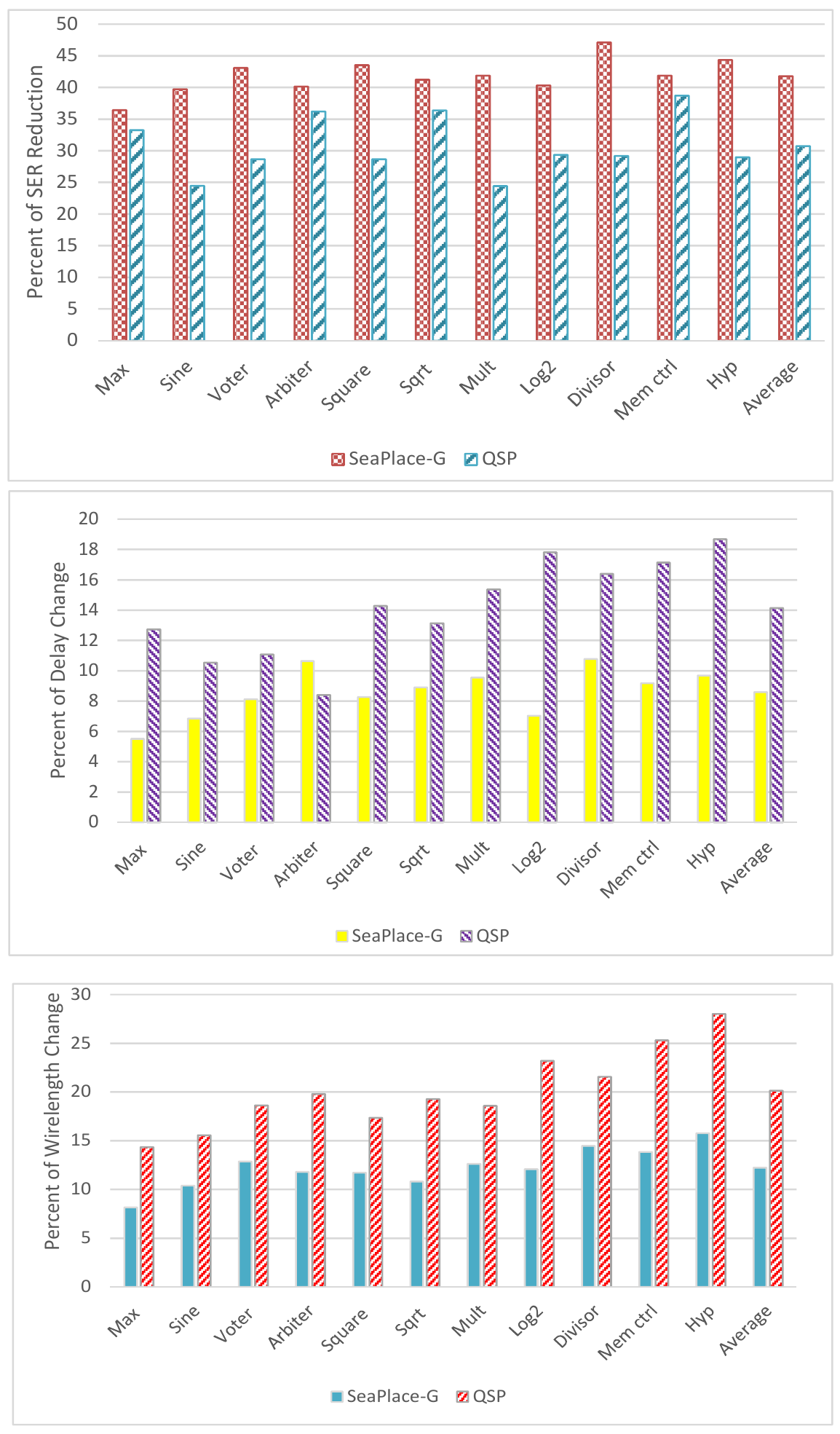}
		\vspace{-0.1in}
		\caption{Comparison of SER Reduction, Delay and Wirelength Overheads of SeaPlace-G and QSP Placements}
		\vspace{-0.2in}
	\label{fig:10}
	\end{center}	
\end{figure}

\begin{figure*}[!tb]
	\begin{center}
		\includegraphics[width = 0.8\textwidth]{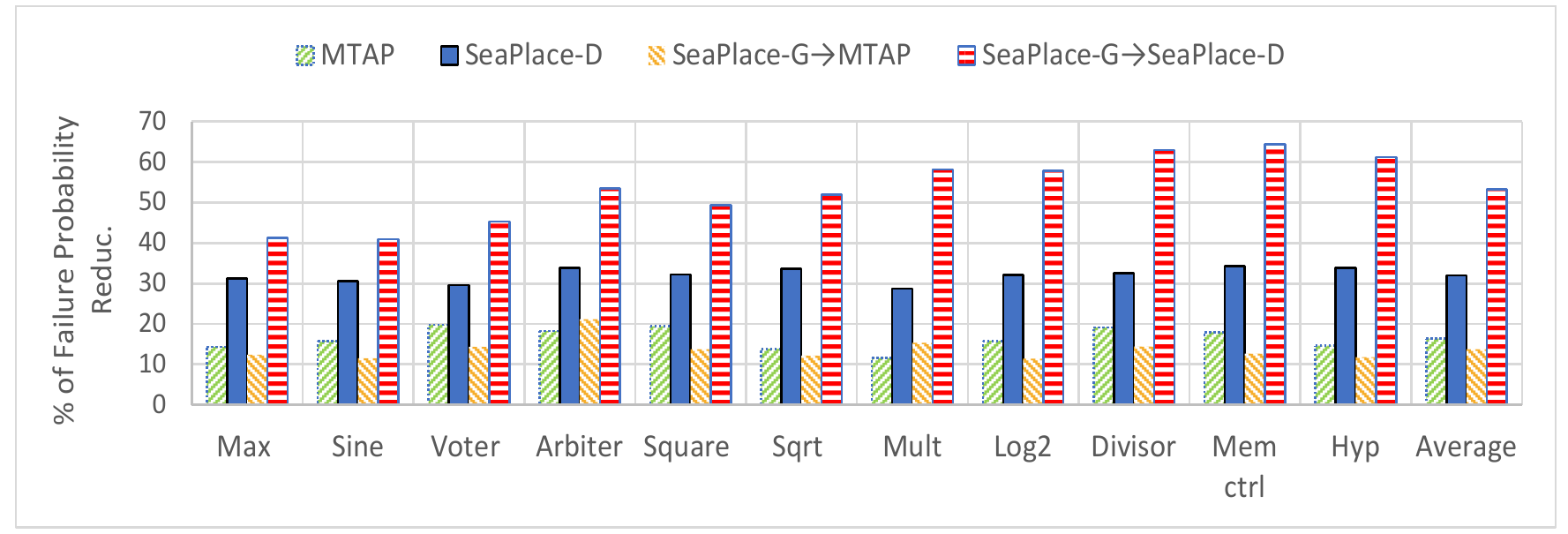}
		\vspace{-0.1in}
	\caption{Failure probability reduction for different combinations of placement (\%).}
		\vspace{-0.2in}
	\label{fig:table2fig}
	\end{center}	
\end{figure*}

\subsection{Comparison to the State-of-the-art Studies}
To assess the efficiency of SeaPlace-G algorithm, its results is compared with an SET-aware quadratic placement referred as QSP \cite{23}, which, as the best we know, is the sole SET-aware global placement in the literature considered three factors of transient fault masking (logical, electrical and timing masking).

Figure \ref{fig:10} presents the results of SeaPlace-G and QSP in terms of SET-originated SER reduction, delay and wirelength overheads. In this experiment, both aforementioned algorithms are compared with the wirelength optimized placement which is obtained using SOC Encounter \cite{44}.

As show in Figure \ref{fig:10}, the average percentage of SER reduction achieved by SeaPlace-G (41.78\%) is greater than that of QSP (30.74\%) and this observation holds for all studied circuits. This is because, SeaPlace-G uses WID variation map information resulting in finding placements with lower SER. Moreover, delay and wirelength overheads of using SeaPlace-G is less than that of QSP for all circuits. This is due to the fact that QSP algorithm increases the length of nets for the sake of using low-pass filtering nature of long nets. Although this approach may lead to SER reduction, itbut imposes high delay and wirelength overheads.

To investigate the efficiency of our SeaPlace-D algorithm, we compare its results with \cite{24} which presents an SER aware detail placement referred as MTAP. In MATP, the location of cells within each row is adjusted by redistribution the existing whitespaces without changing the order of cells in each row.

Figure \ref{fig:table2fig} illustrates the failure probability reduction for different placement combinations. The first two bars are dedicated to the MTAP and SeaPlace-D. The next two bars are hybrid methods in which MTAP and SeaPlace-D are performed after applying SeaPlace-G; i.e., the output placement of SeaPlace-G is used as an initial placement of MTAP and SeaPlace-D. All these four placement combinations use a random initial placement accomplished by SOC Encounter. In each combination, the reported failure probability reduction of each circuit is calculated with respect to the failure probability of initial placement of the combination.


As shown in Figure \ref{fig:table2fig}, the SeaPlace-D algorithm averagely achieves 32.04\% failure probability reduction in comparison to the MTAP with a failure probability reduction of 16.42\%. The reason of this improvement is that, the proposed SeaPlace-D algorithm considers WID process variation information and also uses more different mechanisms to move the cells across multiple rows.

In Figure \ref{fig:table2fig}, it is observed that, the values of failure probability reduction for most of the circuits in SeaPlace-G→MTAP row are lower than their corresponding values in MTAP row. Since the MTAP algorithm redistributes white spaces in each row, it needs lots of whitespaces in each row to be able to adjust the cells’ locations. We observed that, the number of whitespaces in each row is usually reduced after applying SeaPlace-G to the initial placement. As a result, the amount of failure probability reduction for the average and the most of the circuits in SeaPlace-G→MTAP case is fewer than their values in MTAP case. Nevertheless, there are some circuits that, their failure probability reduction values in SeaPlace-G→MTAP row are greater than those of MTAP row. This can be explained by differences in pulse masking ability between these two placement combinations. In other words, SeaPlace-G achieves high amounts of pulse masking and subsequently SET-originated SER reduction for these circuits, as previously presented in Figure \ref{fig:table2fig}. So, these circuits have considerable potential to mask MET-based transient pulses, provided their masking characteristics are not negatively affected by MTAP.

Another observation is that, SeaPlace-G→SeaPlace-D combination has the most average failure probability reduction (53.3\%) among all combinations. In this combination, SeaPlace-G hardens the circuits against SETs by enhancing three masking effects in them. Also, SeaPlace-D not only focuses on hardening against METs by increasing the chance of pulse quenching effects, but also puts a constraint on maximum allowed changes in SET-originated SER of the circuits, i.e., $\Delta SER$. Hence, the mechanisms of hardening against SETs positively affect the MET-aware failure probability reduction of circuits.

\begin{figure}[!tb]
	\begin{center}
		\includegraphics[width = 0.4\textwidth]{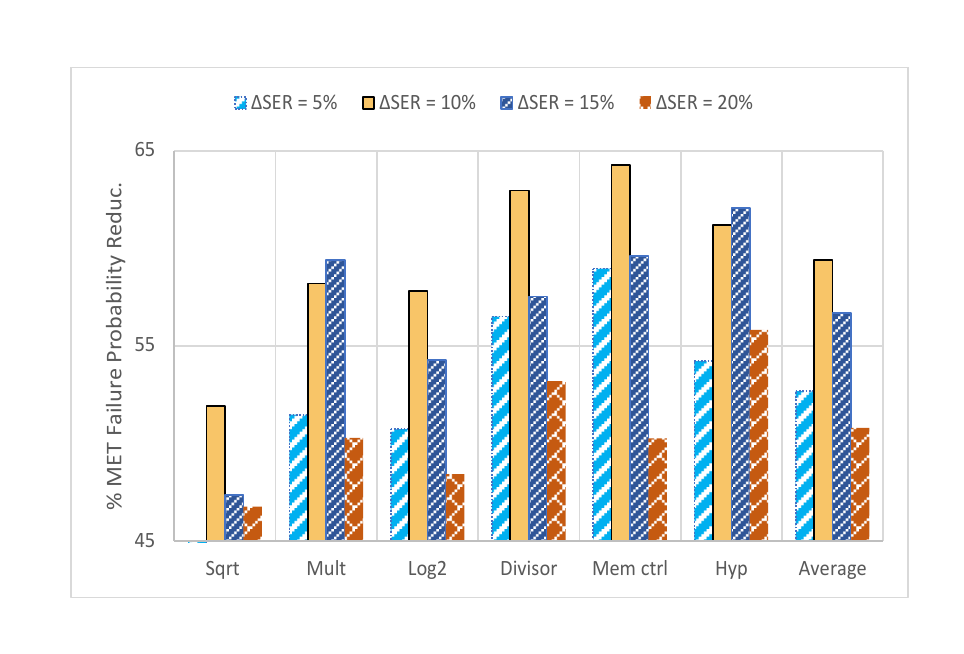}
		\vspace{-0.1in}
		\caption{Percent of MET failure probability reduction achieved by SeaPlace-G$\rightarrow$SeaPlace-D for multiple $\Delta SER$ values}
		\vspace{-0.2in}
	\label{fig:13}
	\end{center}	
\end{figure}

In order to investigate the effect of $\Delta SER$ on failure probability reduction of SeaPlace-D algorithm, an experiment is conducted to compare the achieved failure probability reduction of some circuits while applying different values of $\Delta SER$ to each circuit. Figure \ref{fig:13} shows the percent of MET-originated failure probability reduction for some large circuits of EPFL benchmark circuits in which different values of failure probability reduction for each circuit are obtained under multiple values of $\Delta SER$.

As shown in Figure \ref{fig:13}, for the cases with a $\Delta SER$ lower than 15\%, the average percent of MET failure probability reduction is increased by increase in $\Delta SER$ since further moves are performed, resulting in more pulse quenching effects. Performing even more moves result in more pulse quenching effects, but it may negatively affect the SET-originated SER as the location and subsequently electrical and timing masking factors of moved cells are changed. These changes in pulse masking factors not only change the SET-originated SER but also affect the MET-originated failure probability for the circuits. In Figure \ref{fig:13}, MET failure probability reduction for $\Delta SER=20\%$ is decreased in comparison to  $\Delta SER=15\%$ for all circuits, which is due to a compromise between increasing pulse quenching effects and decreasing masking factors for some cells. Also, the decrease of MET failure probability reduction for Mult and Hyp benchmarks for some amounts of $\Delta SER$ can be explained by this compromise.

\subsection{Runtime Overhead}
\begin{figure}[!tb]
	\begin{center}
		\includegraphics[width = 0.48\textwidth]{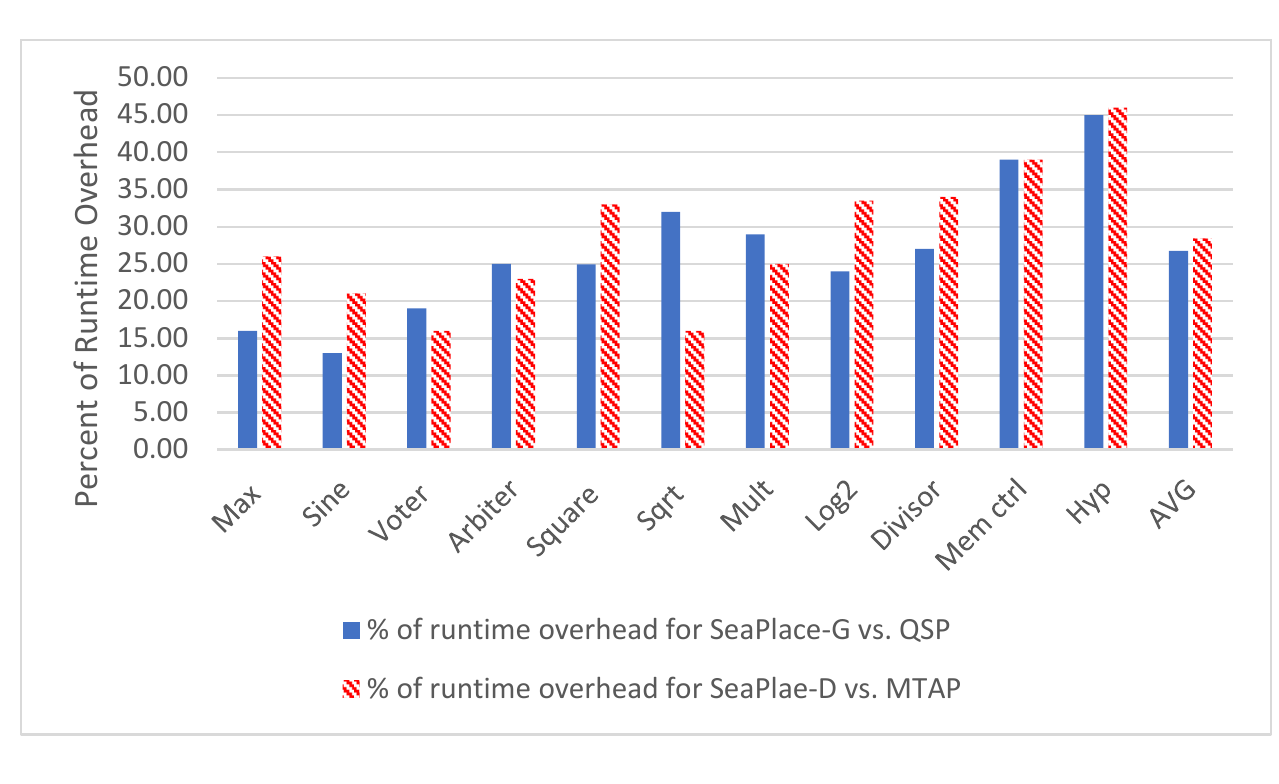}
		\vspace{-0.1in}
	\caption{Percent of runtime overheads for SeaPlace-G vs. QSP  and SeaPlae-D vs. MTAP.}
		\vspace{-0.2in}
	\label{fig:14}
	\end{center}	
\end{figure}

Figure \ref{fig:14} presents the runtime overheads for SeaPlace-G vs. QSP  and SeaPlae-D vs. MTAP. The results show that, the runtime of SeaPlace-G for studied circuits is 26.72\% more than the runtime of QSP, averagely; because SeaPlace-G has more complex objective function and additional constraints to consider WID variation information. Also, the execution of SeaPlace-D placement has 28.41\% runtime overhead in comparison with MTAP method, that is because of solving some consecutive LP optimizations. It should be mentioned that SeaPlace-G and SeaPlace-D incur acceptable runtime overhead in all circuits, and the maximum amount of runtime overhead for SeaPlace-G and SeaPlace-D is about 47 and 13 minutes, respectively.

\section{Conclusion} \label{conclusion}
In this paper, novel WID process variation-aware placement algorithms for improving soft error reliability of combinational circuit against both SETs and METs are presented. Unlike previous placement-based SER reduction methods, in which the effect of WID process variations on their method is neglected, we modeled and utilized WID variation information in our placement algorithms aimed for increasing soft error reliability. Our algorithms included a global placement called as SeaPlace-G and a detailed placement named as  SeaPlace-D.  SeaPlace-G was proposed for increasing circuit reliability against SETs and  SeaPlace-D was presented for hardening against METs. Experimental results showed that on average, SeaPlace-G achieved 41.78\% SER reduction against SETs, SeaPlace-D and  SeaPlace-G→SeaPlace-D (SeaPlace-G followed by the SeaPlace-D) reduced the failure probability against METs by 32.04\% and 53.3\%, respectively.

\bibliographystyle{IEEEtran}
\bibliography{SR_References}
\end{document}